Novel approach for old results on plasma physics: what new do we learn about?

Sebastiano Tosto

ENEA Casaccia, via Anguillarese 301, 00123 Roma, Italy

sebastiano.tosto@enea.it
stosto@inwind.it

Abstract

The present paper proposes a simple model aimed to point out the link between basilar concepts of plasma physics and fundamental principles of quantum mechanics. The model shows in particular that Debye lengths and plasma frequency are actually straightforward consequences of the indistinguishability of identical particles and the exclusion and uncertainty principles.

Introduction.

The plasma physics covers a wide variety of topics that include controlled thermonuclear fusion, evaporation and ionization induced by laser light interaction with matter, plasma phenomena in gas discharges, dynamics of flows of ionized particles in space and atmospheric plasmas; relevant importance have today also the technological applications of plasmas, e.g. for material synthesis, surface processing and thin film deposition. Besides the numerous literature papers on specific aspects of the plasma science, several textbooks highlight the basic physical principles that govern the properties of plasmas, e.g. [1], and the contribution of computer simulation in understanding general theory [2] and phenomena in fusion and astrophysics [3] or in gas discharges [4]. The present paper proposes a simple model that exploits "ab initio" fundamental ideas of quantum mechanics only. The aim is to infer basilar concepts of plasma physics and emphasize the conceptual basis common to all the aforesaid areas of research. The indistinguishability of identical particles and the exclusion and uncertainty principles will show in particular that Debye lengths and plasma frequency are in effect direct consequence of the basic principles of quantum mechanics.

Physical background.

The present model concerns a non-relativistic gas of $n_e$ electrons having mass $m_e$ confined in the linear space range $\Delta\rho = |\Delta\boldsymbol{\rho}|$ at equilibrium temperature $T_e$ in the absence of applied external potential. Regard $\Delta\rho$ as the physical size of a 1D box where the electrons move by effect of their mutual repulsion and thermal kinetic energy; its arbitrary size corresponds thus by definition to the delocalization extent of $n_e$ electrons. The only assumption of the model is $\Delta p_\rho \approx \hbar/\Delta\rho$, where $\Delta p_\rho$ is the range including the momenta components of all the electrons along $\Delta\boldsymbol{\rho}$. No hypotheses are necessary about $\Delta\rho$ and $\Delta p_\rho$. The uncertainty principle prevents knowing local position and momentum of electrons; it is possible however to define their average distance $\overline{\Delta\rho_{ne}} = \Delta\rho/(n_e-1)$ and also to introduce the sub-range $\delta\rho < \Delta\rho$ encompassing the random distance between any local couple of contiguous electrons. Whatever $\delta\rho$ might be, its size must be a function of time in order to contain two electrons moving away each other because of their electric repulsion. To describe the dynamics of this couple, consider first the general problem of two charges $\delta\rho$ apart and let $\delta p_\rho$ be the range including the local momenta components $p_\rho$ allowed by their electric interaction. In general and without any hypothesis $\delta p_\rho$ must have the form $p_\rho - p_{o\rho}$ or $p_{o\rho} - p_\rho$ with $p_\rho$ time dependent and $p_{o\rho}$ constant, both arbitrary; the latter is defined by the momentum reference system, the former by the interaction strength. So $\dot{p}_\rho = \delta \dot{p}_\rho \approx \pm\hbar\delta\dot{\rho}/\delta\rho^2$ is the repulsion/attraction force experienced by one charge by effect of the other. Consider now the upper sign to describe in particular the mutual repulsion between two electrons and make the expansion rate $\delta\dot{\rho}$ tending asymptotically to $c$ to ensure that the electrons cannot travel beyond $\delta\rho$ whatever their current repulsion force might be; this chance, in fact allowed by the arbitrary sizes of $\delta\rho$ and $\Delta\rho$, gives the sought repulsion force $\dot{p}_\rho \approx e^2/(\alpha\delta\rho^2)$, being $\alpha$ the fine structure constant. Introduce now a proportionality constant, $\varepsilon'_0$, to convert the order of magnitude link provided by the uncertainty principle into an equation; merging $\alpha$ and $\varepsilon'_0$ into a unique constant, $\varepsilon_0$, the force between the electrons has the well known form $e^2/(\varepsilon_0\delta\rho^2)$ with $\varepsilon_0$ defined by the charge unit system. The average repulsion energy between any isolated pair of contiguous electrons at distance $\overline{\Delta\rho_{ne}}$, i.e.

neglecting that of all the other electrons, is then $\bar{\eta}_{cont} = e^2 / (\varepsilon_0 \overline{\Delta\rho_{ne}})$; also, the average repulsion energy acting on one test electron by effect of all the others is $\bar{\eta}_{rep} = (n_e - 1)^{-1} \sum_{i=1}^{n_e-1} e^2 / (\varepsilon_0 \Delta r_i)$, where $\Delta r_i$ are the distances between the $i$-th electrons having local coordinates $r_i$ and the test electron. Let us put now by definition $\overline{\Delta\rho_{ne}}^{-1} = f \sum_{i=1}^{n_e-1} \Delta r_i^{-1}$. Formally this equation replaces the sum of all the unknown actual distances of the electrons from the test electron with the reciprocal average distance through the unique arbitrary parameter $f \neq 0$, by definition positive, describing the possible configurations of the electron system; one expects by consequence a simpler expression of $\bar{\eta}_{rep}$ as a function of $\overline{\Delta\rho_{ne}}$ and, through this latter, of $\Delta\rho$ as well. The condition that all the electrons must be in $\Delta\rho$ requires $\Delta r_i < \Delta\rho$, i.e. $\Delta r_i = \chi_i \Delta\rho$ with $\chi_i < 1$; so $(n_e - 1)/f = \sum_{i=1}^{n_e-1} \chi_i^{-1}$ is fulfilled with a proper choice of $f$ whatever the various $\Delta r_i$ might be. In turn $\chi_i < 1$ require $f < 1$, since for $f = 1$ all the $\Delta r_i$ should be equal to $\Delta\rho$. Hence

$$\Delta\rho^{-1} = (n_e - 1)^{-1} f \sum_{i=1}^{n_e-1} \Delta r_i^{-1} \qquad 0 < f < 1 \qquad 1$$

Also, expressing $\Delta r_i^{-1}$ as a function of $1/\overline{\Delta\rho_{ne}}$ as $\Delta r_i^{-1} = \xi_i / \overline{\Delta\rho_{ne}}$ through the parameters $\xi_i > 0$ one finds $\sum_{i=1}^{n_e-1} \xi_i = 1/f$; regardless of the unknown and arbitrary sizes of $\Delta\rho$ and $\Delta r_i$, this result simply requires $\chi_i \xi_i = (n_e - 1)^{-1}$. Eqs 1, possible in principle from a mathematical point of view, have also physical interest because they relate $\Delta\rho$ and $\overline{\Delta\rho_{ne}}$ to $\bar{\eta}_{rep}$:

$$\bar{\eta}_{rep} = \frac{e^2}{f \varepsilon_0 \Delta\rho} \qquad\qquad \bar{\eta}_{cont} = \frac{e^2}{\varepsilon_0 \overline{\Delta\rho_{ne}}}$$

In conclusion our degree of knowledge about the system is summarized by $\bar{\eta}_{rep}$ and $\bar{\eta}_{cont}$, linked by the unknown parameter $f$: the former energy concerns the average collective behaviour of all the electrons, the latter that of a couple of electrons only. On the one side, this conclusion is coherent with the general character of the present approach that disregards specific values of local dynamical variables; in effect any kind of information about $f$ would unavoidably require some hypothesis on the conjugate dynamical variables themselves, which are instead assumed completely random, unknown and unpredictable within their respective uncertainty ranges. On the other side, just the impossibility of specifying the various $r_i$, which prevents establishing preferential values of $f$, compels regarding the properties of the electron gas through its whole uncertainty range $\Delta\rho$ and the whole range of values allowed for $f$. In other words, to each value possible for $f$ corresponds a possible electron configuration of the system physically admissible. For instance, consider in particular the chances $f \to 0$ or $f \to (n_e - 1)^{-1}$ or $f \to 1$ to illustrate at increasing values of $f$ the related information about the respective electron configurations. The first chance $f \to 0$ requires at least one or several $\Delta r_i$ tending to zero, because the possibility of finite $\Delta\rho$ cannot be excluded whatever $f$ might be; to this clustering effect around the test electron corresponds thus an expected increase of $\bar{\eta}_{rep}$. Note however that even in this case, in principle possible, the average energy $\bar{\eta}_{cont}$ between any couple of electrons does not diverge being defined by $\overline{\Delta\rho_{ne}}$ only. This result alone describes of course only a partial aspect of the real plasma state; more exhaustive physical information is obtained examining the further choice of values possible for $f$. If $f \to (n_e - 1)^{-1}$ then $\bar{\eta}_{rep} \to \bar{\eta}_{cont}$, i.e. the average repulsion energy $\bar{\eta}_{rep}$ acting on the test electron tends to that of an isolated couple of contiguous electrons. Moreover, if $f \to 1$ then $\xi_i < 1$ mean $\Delta r_i > \overline{\Delta \rho_{ne}}$, i.e. the distances of the various electrons from the test one are greater than the average value; in this case $\overline{\eta}_{rep}$ tends to $\overline{\eta}_{cont}/(n_e - 1)$, i.e. it is even smaller than before. Summarizing the discussion above, the clustering of electrons around the test electron appears energetically unfavourable, whereas more likely result instead increasing values of $f$ that diminish $\overline{\eta}_{rep}$ down to $\overline{\eta}_{cont}$ or to the smaller value $\overline{\eta}_{cont}/(n_e - 1)$ tending even to zero for large $n_e$. So, without specifying any electron in particular and noting that by definition $\overline{\eta}_{cont}$ corresponds to the energy of a test charge $e$ on which act all the other $(n_e - 1)e$ charges located at the maximum possible distance $\Delta \rho$, the whole range of values allowed to $f$ reveals the preferential propensity of the system to create holes in the linear distribution of average charge around any test electron that screen the repulsive effect of other mobile electrons. Since the test electron is indistinguishable with respect to the others, this behaviour holds in fact for any electron; it is essential in this respect the random motion of mobile electrons that tend to repel each other, not $n_e$ or the local position and momentum of each electron. This picture of the system, which holds regardless of the actual size of $\Delta \rho$ and despite the lack of specific values definable for $f$, is further exploited considering that two electrons are allowed in each energy state corresponding to their possible spin states. Whatever the specific value of $f$ might be, let us examine the behaviour of one of such couples with energy $\overline{\eta}_{cont}$ formed by the test electron and one among the $n_e - 1$ residual electrons at distance $\Delta \rho$, assuming in the following $n_e \gg 1$ and $T_e$ high enough to regard the electrons as a classical gas of particles characterized by random motion in agreement with the previous reasoning. Each electron of the test couple has thus average Coulomb energy $\overline{\eta}_{cont}/2 = n_e e^2/(2\varepsilon_0 \Delta \rho)$ and thermal kinetic energy $\overline{\eta}_{th} = 3KT_e/2 + \overline{\eta}_{corr}/2$; the first addend describes the electrons of the couple as if they would be free gas particles, the second is the obvious correction due to their actual electrical interaction. It will be shown below that in fact this correction term can be neglected if the gas is hot enough, because of the propensity of the system to high values of $f$ that reduce $\overline{\eta}_{rep}$; yet we consider here for generality both terms, noting that $\overline{\eta}_{corr}$ is an unknown function of $n_e$ and $f$ (more exactly of $f^{-1} - 1$) since it depends on the shielding strength between the electrons of the couple provided by the charge holes within $\Delta \rho$. The factor 3, which accounts for three freedom degrees of thermal motion, does not conflict with the electron confinement within the linear delocalization range $\Delta \rho$. The previous discussion has regarded the size of uncertainty range only, rather than the vector $\Delta \boldsymbol{\rho}$ that is actually not uniquely defined: owing to the lack of hypotheses about $\Delta \boldsymbol{\rho}$, any vector with equal modulus randomly oriented with respect to an arbitrary reference system is in principle consistent with the aforesaid results. So there is no reason to think $\Delta \boldsymbol{\rho}$, whatever its actual modulus might be, distinctively oriented along a prefixed direction of the space; thus cannot be excluded even the idea that the orientation of $\Delta \boldsymbol{\rho}$ changes as a function of time. In fact this conclusion suggests that, under proper boundary conditions, the whole vector $\Delta \boldsymbol{\rho}$ can be considered free to rotate randomly in the space at constant angular rate $\boldsymbol{\omega}'$. This simply means introducing into the problem the angular position uncertainty of the electrons together with their radial uncertainty, the only one so far concerned. The simultaneous angular motion of all the electrons does not change the reasoning above about the mutual repulsion energies, while any possible alteration of electron configuration in the rotating frame is still described by the parameter $f$ in its unchanged range of values: nothing was known about the possible local electron configurations before introducing the frame rotation, nothing is known even now about their possible modification. Although the space orientation of the rotation axis is clearly indefinite, introducing the angular uncertainty helps to explain the physical meaning of average energy $\bar{\eta}_{cont}$: on average, the $n_e - 1$ electrons moving randomly in radial direction are statistically distributed on the surface of a sphere of radius $\Delta\rho$ centred on the test electron. Plays a crucial role in this context the indistinguishability: this picture holds for any electron, without possibility of specifying which one, and thus in fact for all the electrons of the system. The lack of further information does not preclude however to define the total energy balance of the test couple electron delocalized within $\Delta\rho$ with velocity resulting by: (i) its momentum randomly falling within the range $\Delta p_\rho$, (ii) the angular motion of its delocalization range $\Delta\boldsymbol{\rho}$ as a whole and (iii) the thermal random contribution, whose average modulus is $v_{th}^2 = (3KT_e + \bar{\eta}_{corr})/m_e$. Of course the modulus $\omega'$ is not arbitrary: its value is in fact determined by the driving energies of the system because the angular, thermal and electric terms must fulfil the condition $\bar{\eta}_{ang} = \bar{\eta}_{th} + \bar{\eta}_{cont}/2$. In this way the energy of angular motion of the test electron described by $\bar{\eta}_{cont}$ is the same as that in the non-rotating linear range $\Delta\rho$, but simply expressed in a different form, i.e. as a function of the angular uncertainty instead of the linear uncertainty only. Moreover, whatever the random motion of each electron might be, the concurrence of radial and angular uncertainties agrees with the energy conservation expected for the whole isolated system described by average quantities only. Wherever the current position of the test electron in the space might be, the average angular motion energy reads $\bar{\eta}_{ang} = m_e \omega'^2 \Delta\rho^2/2$: since $\bar{\eta}_{cont}$ appearing at right hand side of the energy balance is the average energy of the concerned electron couple, the aforementioned reasoning suggests to regard the test electron at distance $\Delta\rho$ from the other $n_e - 1$ electrons around it. Then

$$\frac{m_e \omega'^2 \Delta\rho^2}{2} = \frac{3}{2} KT_e + \frac{\bar{\eta}_{corr}}{2} + \frac{n_e e^2}{2\varepsilon_0 \Delta\rho} \qquad 2$$

As expected, this result does not depend on some particular value of $f$ in the approximation of negligible $\bar{\eta}_{corr}$ and gives

$$\omega'^2 - \frac{\bar{\eta}_{corr}}{m_e \Delta\rho^2} = 3\frac{KT_e}{m_e \Delta\rho^2} + \frac{n_e e^2}{m_e \varepsilon_0 \Delta\rho^3}$$

Introduce now the electron number density $N_e = n_e/\Delta\rho^3$ and define the correction term according to its physical dimensions putting $\bar{\eta}_{corr}/(m_e \Delta\rho^2) = \omega_{corr}^2$. Moreover exploit the fact that the test electron rotating along the circumference $2\pi\Delta\rho$ has De Broglie's wavelength $\lambda = 2\pi\Delta\rho$ and momentum $p_\lambda = \hbar k$, being $k = 2\pi/\lambda$. Here has been considered the fundamental oscillation only, omitting the shorter wavelengths described by integer multiples $n\lambda = 2\pi\Delta\rho$ of $\lambda$. One finds thus

$$\omega^2 = \frac{N_e e^2}{\varepsilon_0 m_e} + 3k^2 \frac{KT_e}{m_e} \qquad N_e = \frac{n_e}{\Delta\rho^3} \qquad \omega^2 = \omega'^2 - \omega_{corr}^2 \qquad \omega_{corr}^2 = \frac{\bar{\eta}_{corr}}{m_e \Delta\rho^2} \qquad k = \frac{1}{\Delta\rho} \qquad 3$$

This result assigns to the frequency $\omega$ the physical meaning of collective property of electrons, owing to the fact that it is defined through average quantities. In this result is hidden the electron characteristic length $\lambda_{eD}$ as well; the first eq 3 can be indeed rewritten more expressively as follows

$$\omega^2 = \omega_p^2 (1 + 3k^2 \lambda_{eD}^2) \qquad \omega_p^2 = \frac{N_e e^2}{\varepsilon_0 m_e} \qquad \lambda_{eD} = \sqrt{\frac{\varepsilon_0 KT_e}{e^2 N_e}}$$

This result, well known, can be further exploited considering fixed $\omega^2$ in eq 3. Eq 2 shows that for $\Delta\rho$ large enough the Coulomb term becomes negligible with respect to the thermal energy, in which case $\Delta\rho_{eD}^2 \approx 3KT_e/(m_e\omega^2)$; the subscript denotes the particular value of $\Delta\rho^2$ fulfilling this limit condition of the test electron in the gas, which justifies why $\bar{\eta}_{th}$ is in fact well approximated by the free electron energy term only, i.e. $\omega^2 \approx \omega'^2$. Replace now $m_e \omega^2$ in the first eq 3 regarded in particular for $\Delta \rho^2 \equiv \Delta \rho_{eD}^2$ and $\Delta p_\rho^2 \equiv \Delta p_{eD}^2$; then $\Delta \rho_{eD}^2 = (N_e e^2/(3KT_e \varepsilon_0) + \Delta p_{eD}^2/\hbar^2)^{-1}$. In general, large $\Delta \rho$ entails accordingly small $\Delta p_\rho/\hbar$. If $\Delta p_{eD}/\hbar$ is small enough and $N_e$ high enough to have $N_e e^2/\varepsilon_0 >> 3KT_e \Delta p_{eD}^2/\hbar^2$ the result is $\Delta \rho_{eD}^2 \rightarrow 3KT_e \varepsilon_0/(N_e e^2) = 3\lambda_{eD}^2$. Clearly the vanishing Coulomb term means that the screening effect due to the motion of the plasma charges is controlled by the characteristic scale length $\lambda_{eD}$. At this point it is also immediate to infer what changes if instead of an electron gas only one considers a plasma made by $n_e$ electrons at average temperature $T_e$ plus $n_i$ ions with charge $-Ze$ at average temperature $T_i$. Consider first in $\Delta \rho$ the ion gas only. Simply repeating the reasoning above, the result becomes $\Delta \rho_{iD}^2 = (N_i Z^2 e^2/(3KT_i \varepsilon_0) + \Delta p_{iD}^2/\hbar^2)^{-1}$, analogous to that obtained before; of course also now $\Delta \rho_{iD}$ is obtained through the positions $\Delta \rho^2 \equiv \Delta \rho_{iD}^2$ and $\Delta p_\rho^2 \equiv \Delta p_{iD}^2$. If ions and electrons are both confined in the same range, $\Delta \rho_{eD}$ and $\Delta \rho_{iD}$ must coincide. Observing the results just obtained, this condition appears fulfilled if $\Delta p_{iD}^2/\hbar^2 = N_e e^2/(3KT_e \varepsilon_0)$ and $\Delta p_{eD}^2/\hbar^2 = N_i Z^2 e^2/(3KT_i \varepsilon_0)$; then, the position $\Delta p_D^2 = \Delta p_{eD}^2 + \Delta p_{iD}^2$ gives $\Delta \rho_D^2 = \hbar^2 (\Delta p_{eD}^2 + \Delta p_{iD}^2)^{-1} = \hbar^2 \Delta p_D^{-2}$, as expected. The electro-neutrality $N_p = N_e = N_i Z$ gives the well known result

$$\Delta \rho_D^2 = 3\lambda_D^2 \qquad \lambda_D^2 = \frac{K\varepsilon_0}{e^2 N_p} \frac{1}{T_e^{-1} + ZT_i^{-1}} \qquad N_p = N_e = N_i Z \qquad \Delta \rho_D \equiv \Delta \rho_{eD} \equiv \Delta \rho_{iD} \qquad 4$$

Discussion.
Eqs 3 and 4 evidence that the expression of $\lambda_D$, the plasma frequency and the basic concepts of plasma physics are simply hidden within the quantum uncertainty, from which they can be extracted through an elementary and straightforward reasoning. For comparison purposes, it is instructive at this point to remind shortly the two key-steps through which are usually inferred the plasma properties: (i) to assume Coulomb law and Boltzmann-like number density of electrons/ions, according to the idea that a high local probability of finding a particle is related to a high local charge density; (ii) to solve the potential Poisson equation assuming $eV << KT_e$, being $V$ the electric potential. The present approach is of course necessarily different, because the lack of information about the distances between the electrons compels introducing the parameter $f$ that replaces the unknown quantities $\Delta r_i$ and because the plasma properties are inferred through eqs 1 containing average quantities only; moreover the classical assumption $\overline{\eta}_{th} \approx 3KT_e/2$ replaces $eV << KT_e$. Strictly speaking $f$ cannot be considered unknown, since in fact it is conceptually not definable by assigned values; its arbitrariness is nothing else but that physically inherent the uncertainty ranges $\Delta \rho$ and $\Delta p_\rho$. Yet just $\Delta \rho$ enabled the well known Debye lengths of electrons and ions to be also found through the simple energy balance of the respective test charges and the boundary condition of electro-neutrality of the plasma. Defining the characteristic lengths $\Delta \rho_{eD}$ and $\Delta \rho_{iD}$ does not mean however determining the range $\Delta \rho$, which remains arbitrary because of the presence of the terms $\Delta p_{eD}^2/\hbar^2$ and $\Delta p_{iD}^2/\hbar^2$. Note in this respect that the plasma properties require $\Delta \rho > \Delta \rho_D$, i.e. they appear when the delocalization extent of the gas of charged particles is greater than the Debye length of electrons and ions, which fix therefore the length scale above which hold the peculiar features of plasma physics. Replacing momenta and coordinates with the respective uncertainty ranges means renouncing "a priori" to any information about motion and position of particles; consequently no specific representation of the electron system could have been expected through such a conceptual background. Consider for instance the approximate solution of Poisson's equation $V \sim r^{-1}\exp(-r/\lambda_D)$, calculable as a function of $r$; here of course such an information is missing once having skipped $r$. Yet, on the one side the information inferred about the system is enough to highlight the same physical consequences, e.g. the reduced penetration depth of the electric field within the plasma; on the other side, once having found the Debye lengths that control this depth, one could easily infer the quoted form of $V$ simply repeating backwards with the help of Coulomb law and Boltzmann's distribution the well known mathematical steps. Doing so however would not add any crucial contribution to the previous considerations based on $f$ only, apart the necessity of regarding such $r$ as a parameter significant within a few Debye lengths but not exactly defined point by point like a classical coordinate. Nevertheless the present approach, apparently more agnostic, enabled also the Coulomb law to be inferred itself. Moreover further information is easily inferred from eq 2. Consider for simplicity an electron gas only and the energy $\hbar\omega_p$ corresponding to the plasma frequency already calculated consistently with the average energy balance $\bar{\eta}_{cont}$ of any couple; trivial manipulations give

$$\hbar\omega_p = 2\mu_B H \qquad \mu_B = \frac{e\hbar}{2m_e} \qquad H = \sqrt{\mu_0 dc^2} \qquad \mu_0\varepsilon_0 = c^2 \qquad d = \frac{n_e m_e}{\Delta\rho^3} \qquad 5$$

Here $\mu_B$ is the Bohr magneton, while $dc^2$ is the electron rest mass energy density per unit volume corresponding to the number density $N_e$; elementary dimensional considerations show that the third equation defines a magnetic field. Formally the first equation 5 is a possible way to rewrite the acknowledged expression of $\omega_p$ through well known positions, whereas $H$ can be in principle understood because any moving charge generates a magnetic field; yet its true physical meaning would be unclear without knowing that $\omega_p$ has been introduced to describe a couple of electrons with the same average energy $\bar{\eta}_{cont}$, thus with anti-aligned spins, moving away each other and rotating solidly with $\Delta\boldsymbol{\rho}$ as well. In effect the test electron generates a magnetic field normal to its rotation plane; so the direction of $H$ is defined in the reference system of a single couple, whereas its macroscopic average cancels out both because motion and orientation of several mobile electron couples are random and because the rotation axis defining $\omega$ is not uniquely defined itself. Thus $-\mu_B H$ and $\mu_B H$ are the energies of magnetic dipole moment of the electrons of the couple with spin components necessarily opposite with respect to the direction of their own local field; $\hbar\omega_p$ calculates the spin flip energy gap along the local $H$, i.e. the excitation energy $\Delta\bar{\eta}_{cont}$ of the couple. This confirms that the plasma frequency is really a property of any local couple rotating at angular rate $\omega_p$, yet without contradicting the definition of collective property previously assigned to the plasma frequency: of course the couples are not rigidly formed by specific electrons, rather they involve any neighbours randomly approaching or moving away each other. Otherwise stated, owing to the statistical concept of average, $\bar{\eta}_{cont}$ represents in fact the totality of couples possible in the plasma. Thus the simple inspection of $\hbar\omega_p$ compels regarding the collective properties of plasma as due to two-body interactions between continuously exchanging electrons, whose fingerprint is just the form of eqs 5; remains however intriguing the fact that the local field $H$ results defined through the rest mass energy density of all the electrons. It is clear now why the global behaviour of the charges is described by the full range of values of the parameter $f$, and not by some specific values previously exemplified just to check the kind of information provided by the present reasoning: the key energy controlling the properties of plasma is $\bar{\eta}_{cont}$ that does not depend on $f$. So the plasma can be effectively regarded as combination of all the electron configurations physically possible whatever the respective $\bar{\eta}_{rep}$ might be. Also note that the first eq 5 should have been more properly written as $\hbar\omega_p = \pm 2\mu_B H$: the negative sign of $\sqrt{\omega_p^2}$, to be excluded in a classical context exploiting

Coulomb's law and Boltzmann's statistics only, appears natural in the present quantum-mechanical context that admits a negative energy state $-\hbar\omega_p$ of plasma electrons, i.e. a positron plasma, whereas actually $\mu_B = \pm |e|\hbar/2m_e$; in effect all the reasoning so far carried out would remain unchanged considering a positron gas instead of an electron gas, with the factor 2 still accounting for the same energy gap of spin alignment with respect to $H$. A closing remark helps to better clarify the physical meaning of $p_\lambda$. The first eq 3 written in the form $\omega^2 = \omega_p^2 + k^2 V^2$, with $V^2 = v_{th}^2 - \bar{\eta}_{corr}/m_e = 3KT_e/m_e$, shows that $(\hbar\omega)^2$ exceeds the characteristic plasma energy $(\hbar\omega_p)^2$ by $\delta\bar{\eta}^2 = \hbar^2 k^2 V^2$. The fact that $k \to 0$ entails $\omega^2 \to \omega_p^2$ for $\Delta\rho \to \infty$, very large plasma size, suggests regarding $\delta\bar{\eta}^2$ as a local perturbation of $\omega_p^2$ in an arbitrary point encompassed by $\Delta\rho$; so the plasma oscillation deviates from $\omega_p$ only locally, of course without possibility of specifying where exactly. Thus $\delta\bar{\eta}$ could be a spontaneous quantum fluctuation or the consequence of a transient energy local input injected into the plasma: in any case the perturbation does not propagate to infinity because its lifetime is of the order of $\hbar/\delta\bar{\eta} = (kV)^{-1}$. This means that if for instance on the system of electrons acts a flash of thermal energy that rises the local temperature of the gas, then the electron system reacts and tends again to its natural oscillation frequency $\omega_p$, the only effect of the perturbation energy input being an increased average temperature and related Debye's length. To better explain this way of regarding eqs 3, rewrite identically $\delta\bar{\eta}^2 = \hbar^2 k_w^2 (wV)^2$ with $k_w = 2\pi/\lambda_w$ and $\lambda_w = w\Delta\rho$, so that the dimensionless arbitrary parameter $w$ does not affect $\omega^2$; eqs 3 calculated with $w=1$ and $w>1$ give $3\omega_p^2 (k^2 \lambda_{eD}^2 - k_w^2 \lambda_{weD}^2) = 0$, i.e. $\Delta T_e / T_e = w^2 - 1$. Then the random average velocity of the electron gas increases while the momentum $p_\lambda = \hbar k$ decreases to $p_{w\lambda} = \hbar k_w$; moreover the group velocity $V_g = kV^2/\omega$ of the circulating electron wave decreases with $k$, i.e. the perturbation energy $\delta\bar{\eta}$ is dissipated in a range of the order of $V/\omega$. So one infers: (i) the momentum decrease from $p_\lambda$ to $p_{w\lambda}$ describes an electron wave circulating along a circumference of radius $\Delta\rho$ that attenuates when the radius expands to $w\Delta\rho$ along with the related wavelength increase; (ii) the local electron wave of frequency $\omega$ is necessarily longitudinal, since propagation direction and electric field oscillation are by definition both in the radial rotation plane of $\Delta\rho$. Note that $V/\omega < V/\omega_p$ and that the right hand side ratio is nothing else but $\sqrt{3}\lambda_{eD}$; thus the perturbation wave extinguishes in a range of the order of Debye's length, which clarifies why $\omega^2 \to \omega_p^2$ for $3k^2 \lambda_{eD}^2 \ll 1$. Also note that $\Delta\rho \to \infty$ concerns a longitudinal plane wave for which holds Faraday's law $\mathbf{k} \times \mathbf{E} = \omega \mathbf{H}$; recalling that in this limit $\omega \to \omega_p$ and that the local magnetic field $H$ already found is normal to the rotation plane of the test electron at distance $\Delta\rho$ for the other $n_e$ electrons, the electric field acting on the test electron calculated with the help of eqs 3 and 5 has the sensible form $E = n_e e/(\varepsilon_0 \varphi \Delta\rho^2)$, where $\varphi = \sin(\widehat{\mathbf{kE}})$. In principle therefore $E$ depends on how are mutually oriented $\mathbf{k}$ and $\mathbf{E}$, because $\varphi \to 1$ if the vectors tend to $\mathbf{k} \perp \mathbf{E}$ or $\varphi \to 0$ if the vectors tend to $\mathbf{k} \parallel \mathbf{E}$: in the former case $E$ tends surely to zero, in the latter case the limit $\varphi \Delta\rho^2$ for $\Delta\rho \to \infty$ is undetermined, i.e. $E$ could be zero or infinite depending on the rate with which $\varphi \to 0$ for $\Delta\rho \to \infty$. These limits are particularly interesting as they link this result involving directly $E$ to what we have already discussed about the energy of the test electron via $\bar{\eta}_{rep}$ as a function of the values allowed to $f$: (i) the divergent values of $\bar{\eta}_{rep}$ for $f \to 0$ correspond to the chance $\varphi \Delta\rho^2 \to 0$, (ii) the result $\bar{\eta}_{rep} \to \bar{\eta}_{cont}$ for $f \to (n_e - 1)^{-1}$ corresponds to $\varphi \to 1$ in which case $E$ tends to the expected form $n_e e/(\varepsilon_0 \Delta\rho^2)$ acting on the test electron in the field of the other $n_e$ all at distance $\Delta\rho$, (iii) $f \to 1$ corresponds to $E \to 0$, i.e. $\varphi\Delta\rho^2 \to \infty$ whatever $\varphi$ might be. Although obtained in the particular case of a plane electron wave, this result suggests a conceptual link between $f$ and $\varphi$: the physical meaning of $f$ is thus related to the coupling strength between electric field within $\Delta\rho$ and wave vector of the electron circulating along the circumference $2\pi\Delta\rho$. The possible alignments of **k** and **E** are very easily explained considering that while $\Delta\rho$ rotates the electrons move randomly within $\Delta\rho$; so the combination of radial and angular motion does not produce in general a circular path, which would require instead an electron position fixed somewhere within $\Delta\rho$. This reasoning does not contradict the positions of eqs 2, which concern average quantities only; here instead we are attempting to describe through $f$ or $\varphi$ the local electron configuration, whose detailed knowledge is however forbidden by the quantum uncertainty. Otherwise stated, the radial and angular uncertainties that prevent knowing how are specifically oriented **k** and **E** also prevent knowing how change $f$ and $\varphi$. Since both these parameters are unpredictable and random, one cannot expect a functional relation between them; their link is merely conceptual, i.e. both express the lack of local information about position and momentum of the electrons. Nevertheless, just the fact that eqs 2 agree with the experience, confirms that considering average quantities only gives correct results even disregarding since the beginning any local information. So far $w$ has been not yet specified. Consider the particular value such that $w^2(v_{th}^2 - \bar{\eta}_{corr}/m_e) = c^2$; hence $\omega^2 = \omega_p^2 + k^2 c^2$ whatever $v_{th}$ and $\bar{\eta}_{corr}$ might be. Replacing $v_{th}^2 - \bar{\eta}_{corr}/m_e$ with $c^2$ means that the dispersion relation concerns now the propagation of a transverse electromagnetic wave of frequency $\omega$ travelling in the plasma rather than a matter wave; this holds however for k real, i.e. $\omega^2 > \omega_p^2$, otherwise the wave is attenuated. Hence transverse, longitudinal or mixed waves can propagate in the plasma. The overall conclusion is at this point that it is not necessary to introduce positions and momenta of each electron to infer the basic physical properties of plasma; any local information can be disregarded conceptually since the beginning, i.e. not as a sort of approximation to simplify some calculation. If properly exploited, the lack of knowledge inherent the quantum delocalization is actually valuable source of information, in fact the only one physically allowed by the quantum mechanics. Just because consequence of the uncertainty only, the above way to infer some basic concepts of plasma physics is not trivial duplicate of other well known procedures. Note in this respect that the present model is not an occasional attempt that concerns specifically the plasma only; rather it is much more general, as described in detail in the papers [5,6] concerning the quantization of angular momentum and harmonic oscillations, as well as the energy levels of hydrogenlike and many-electron atoms/ions and diatomic molecules. In effect the considerations carried out above replicate and extend concepts already introduced in the quoted papers. It is remarkable the fact that a classical electron gas can be described through the same ideas that correctly describe also the quantized energy of bound electrons in the field of one nucleus or two nuclei.

Conclusion.
The concept of uncertainty is not mere restrictive limit to our knowledge, rather it must be regarded essentially as a fundamental law of nature and then source of physical information that, if properly exploited, allows to explain "ab initio" the physical properties of quantum systems.

REFERENCES

1 T.J.M. Boyd e J.J. Sanderson, "The Physics of Plasmas", Cambridge University Press, (2003)
2 C.K. Birdsall, "Plasma physics via computer simulation", McGraw Hill, N.Y., (1985)
3 T. Tajima, "Computational plasma physics", Addison Wesley, California, (1989)
4 N.R. Franklin, "Plasma phenomena in gas discharges", Clarendon Press, Oxford,(1976)
5 S. Tosto, Il Nuovo Cimento, vol. 111 B, N. 2, (1996), pp. 193-215
6 S. Tosto, Il Nuovo Cimento, vol. 18 D, N. 12, (1996), pp. 1363-1394